\documentclass[aps,prd,twocolumn,showpacs,floatfix,preprintnumbers,amsfont,amsmath,amssymb,nofootinbib, superscriptaddress]{revtex4-1}

\usepackage{color}
\usepackage{hyperref}
\usepackage[normalem]{ulem}
\usepackage{lipsum}
\usepackage{url}
\usepackage{float}
\usepackage{ctable}
\usepackage{graphicx}
\usepackage{physics}
\usepackage{mathtools}
\usepackage{bm}
\usepackage{siunitx}

\newcommand{\liang}[1]{}

\newcommand{\sk}[1]{}

\newcommand{\refsec}[1]{Sec.~\ref{sec:#1}}

\newcommand{\tabline}{\specialrule{.1em}{.05em}{.05em}}
\newcommand{\head}[2]{\multicolumn{1}{>{\centering\arraybackslash}p{#1}}{#2}}

\newcommand{\be}{\begin{equation}}
\newcommand{\ee}{\end{equation}}
\newcommand{\ba}{\begin{eqnarray}}
\newcommand{\ea}{\end{eqnarray}}

\allowdisplaybreaks

\begin{document}

\title{A New Search Pipeline for Compact Binary Mergers: Results for Binary Black Holes in the First Observing Run of Advanced LIGO}

\author{Tejaswi Venumadhav}
\email{tejaswi@ias.edu}
\affiliation{\mbox{School of Natural Sciences, Institute for Advanced Study, 1 Einstein Drive, Princeton, NJ 08540, USA}}
\author{Barak Zackay}
\affiliation{\mbox{School of Natural Sciences, Institute for Advanced Study, 1 Einstein Drive, Princeton, NJ 08540, USA}}
\author{Javier Roulet}
\affiliation{\mbox{Department of Physics, Princeton University, Princeton, NJ, 08540, USA}}
\author{Liang Dai}
\affiliation{\mbox{School of Natural Sciences, Institute for Advanced Study, 1 Einstein Drive, Princeton, NJ 08540, USA}}
\author{Matias Zaldarriaga}
\affiliation{\mbox{School of Natural Sciences, Institute for Advanced Study, 1 Einstein Drive, Princeton, NJ 08540, USA}}

\date{\today}


\begin{abstract}

In this paper, we report on the construction of a new and independent pipeline for analyzing the public data from the first observing run of advanced LIGO for mergers of compact binary systems. The pipeline incorporates different techniques and makes independent implementation choices in all its stages including the search design, the method to construct template banks, the automatic routines to detect bad data segments (``glitches") and to insulate good data from them, the procedure to account for the non-stationary nature of the detector noise, the signal-quality vetoes at the single-detector level and the methods to combine results from multiple detectors. Our pipeline enabled us to identify a new binary black-hole merger GW151216 in the public LIGO data.
This paper serves as a bird's eye view of the pipeline's important stages.
Full details and derivations underlying the various stages will appear in accompanying papers.
\end{abstract}


\maketitle



\section{Introduction}

The LIGO and Virgo observatories reported the detection of several gravitational wave (GW) events from compact binary coalescence in their First and Second Observing Runs (O1 and O2 respectively)~\cite{LIGOScientific:2018mvr}.
These detections required technically sophisticated analysis pipelines to reduce the strain data. 
This is because typical events are buried under the detector noise, and cannot be simply ``seen"  in raw data at current sensitivities.
Hence, any search for signals in the data needs to properly and precisely model the detector noise.

The simplest model is that the detector noise is stationary and Gaussian in nature. 
Under these assumptions, the best method to detect signals is matched-filtering, which involves creating a bank of possible signals, constructing optimal filters (or templates) for the signals given the noise model, and running the templates over the data. 
The resulting scores are distributed according to known (chi-squared) distributions in the presence or absence of real signals~\cite{Jaranowski2012}.

Unfortunately, both the assumptions underlying matched-filtering fail at some level: the noise statistics vary even on the timescales of the (putative) signals, and there are intermittent non-astrophysical artifacts which are clearly not produced by Gaussian random noise (``glitches") \cite{BlipGlitches}, examples of such disturbances can be found in Ref.~\cite{GravitySpy}. 
These systematics pollute the distribution of the matched-filtering scores. 
Moreover, the templates describing different astrophysical signals have finite overlaps, and thus often trigger on the same underlying noise transients.
Detectable real events lie in the tails of the score distribution, and hence it is crucial to properly correct for systematics in order to maximize the sensitivity to GW events, and to quote reliable false-alarm rates (FARs). 

The official LVC catalog of GW events comprises candidates from two independent pipelines: \texttt{PyCBC}~\cite{2016CQGra..33u5004U} and \texttt{GstLAL}~\cite{PhysRevD.95.042001}.
Additional analysis of the data was presented in Ref.~\cite{NitzCatalog}. 
Each of these pipelines has developed solutions for the data complexities described above. 
In this paper, we describe a new and independent analysis pipeline that we have developed for analyzing the publicly available data from the first observing run of advanced LIGO \cite{GWOSC}. Our solutions and implementation choices were guided by the desire to attain, as much as possible, the ideal of the distributions in the Gaussian case, which are easily understood and interpreted.


First, we developed a method to construct template banks that enumerates not over physical waveforms, but over linear combinations of a complete set of basis functions for their phases. Correlations between templates have a uniform and isotropic metric in this space.

Second, when dealing with systematics, we use procedures with analytically tractable behavior in the case of Gaussian random noise, which enables us to set thresholds based on well-defined probabilities. We developed a simple method to empirically correct for the non-stationary nature of the detector noise (PSD drift). Under this procedure, segments of data with no apparent glitches produce trigger scores with perfect chi-squared distributions.
At the first pass, we attempt to veto out residual ``glitches" using a collection of simple tests (either at the signal-processing level or after triggering), while still using the matched-filtering scores as the ranking statistics to leave the Gaussian ``floor" untouched. 
We also developed methods to condition masked data in a way that guarantees that the following matched filtering step would have zero response to the masked data segments.

Finally, we estimate the background of coincident triggers between the two detectors using time slides (akin to \texttt{PyCBC}). 
Our pipeline includes methods to use the information from background triggers to combine physical triggers from different detectors in a statistically optimal manner for distinguishing astrophysical events from noise transients. 

Our paper is organized as follows: Section~\ref{sec:stages} provides an overview of the stages in the pipeline. 
Section~\ref{sec:descriptions} expands upon each of the stages while omitting derivations and precise details, which we present in accompanying papers~\cite{templatebankpaper, psddriftpaper, vetopaper}. In Section~\ref{sec:search_results} we present the results of our search for binary black hole mergers in O1.

\section{Pipeline stages \label{sec:stages}}

We construct our pipeline in several stages, which are organized as follows:
\begin{enumerate}
    \item \textbf{Construction of a template bank:} We divide the mergers into banks with logarithmic spacing in the chirp mass, and analyze each bank separately. Section~\ref{sec:templatebank} provides further details on the underlying method, and the properties of the resulting banks.
    \item \textbf{Analysis of single detector data:} We first analyze the data streams from the Hanford (H1) and Livingston (L1) detectors separately, as follows:
    \begin{enumerate}
        \item We preprocess data from each detector in chunks of $\simeq \SI{4096}{\second}$. Section~\ref{sec:preprocessing} details our initial signal processing.
        \item We iteratively whiten the data stream, perform several tests to detect and remove bad data segments (``glitches"), and condition the remaining data to preserve astrophysical signals. Sections~\ref{sec:badsegments} and \ref{sec:holefilling} describe this procedure. 
        \item We correct for the non-stationary nature of the noise (PSD drift), which if untreated, systematically pollutes the connection between the matched-filtering scores and probability. Section~\ref{sec:psd_drift} provides more details.
        \item We generate matched-filtering overlaps for the waveforms in our banks with the whitened data stream, apply the PSD drift correction, and record triggers whose matched-filtering scores are above a chosen threshold (Section~\ref{sec:matchedfiltering}).
    \end{enumerate}
    \item \textbf{Coincidence analysis between detectors:} We analyze triggers that are coincident in H1 and L1. In Section~\ref{sec:coincidence}, we describe how we collect coincident triggers with combined incoherent score above a threshold, at both physical (candidates) and unphysical (background) time delays.
    \item \textbf{Refining on a fine grid:} We refine the parameters of the candidates and the background on a finer grid around the triggers in order to account for template bank inefficiency, and allow room for more stringent signal quality vetoes.
    \item \textbf{Trigger quality vetoes:} We apply vetoes on the triggers based on the signal quality, as well as the  data quality. The vetoes have to be applied either at the single-detector level, to avoid biasing the calculation of the coincident background using time slides. Section~\ref{sec:vetoes} lists the vetoes we applied to the triggers.
    \item \textbf{Estimating the significance of candidates:} We use the set of background triggers to estimate the FAR for the candidates at physical lags between H1 and L1. We do this in two stages:
        \begin{enumerate}
            \item We first compute a ranking score that is purely a function of the incoherent scores of the triggers, under the assumption that the noise processes that produce the background are independent between detectors (Section~\ref{sec:incoherent}).
            \item Section~\ref{sec:coherent} describes our coherent score, which adds all the information encapsulated in the phase, amplitude, relative sensitivity and arrival time differences between the detectors to create our final candidate ranking statistic.
            \item Section~\ref{sec:pastro} describes how we construct an estimate for the probability of a coincident event being of astrophysical origin given an astrophysical event rate.
        \end{enumerate}
\end{enumerate}



\section{Concise description of the pipeline stages}
\label{sec:descriptions}

\subsection{Template bank}
\label{sec:templatebank}

We perform our search by matching the strain data to a discrete set of waveform templates that sufficiently closely resemble any gravitational wave signal within our target parameter space.
We target our search at coalescing binary black holes (BBH), defined here as compact binary objects with individual masses between $3\,M_\odot$ and $100\,M_\odot$ and with aligned spins. We allow spin magnitudes up to $\abs{\chi_{1,2}} < 0.85$. We restrict the mass ratios to be $q^{-1}<18$.

\begin{figure}
    \centering
    \textbf{\,\,\,\,\,\, Template banks used in the search}\par\medskip    
    \includegraphics[width=\linewidth]{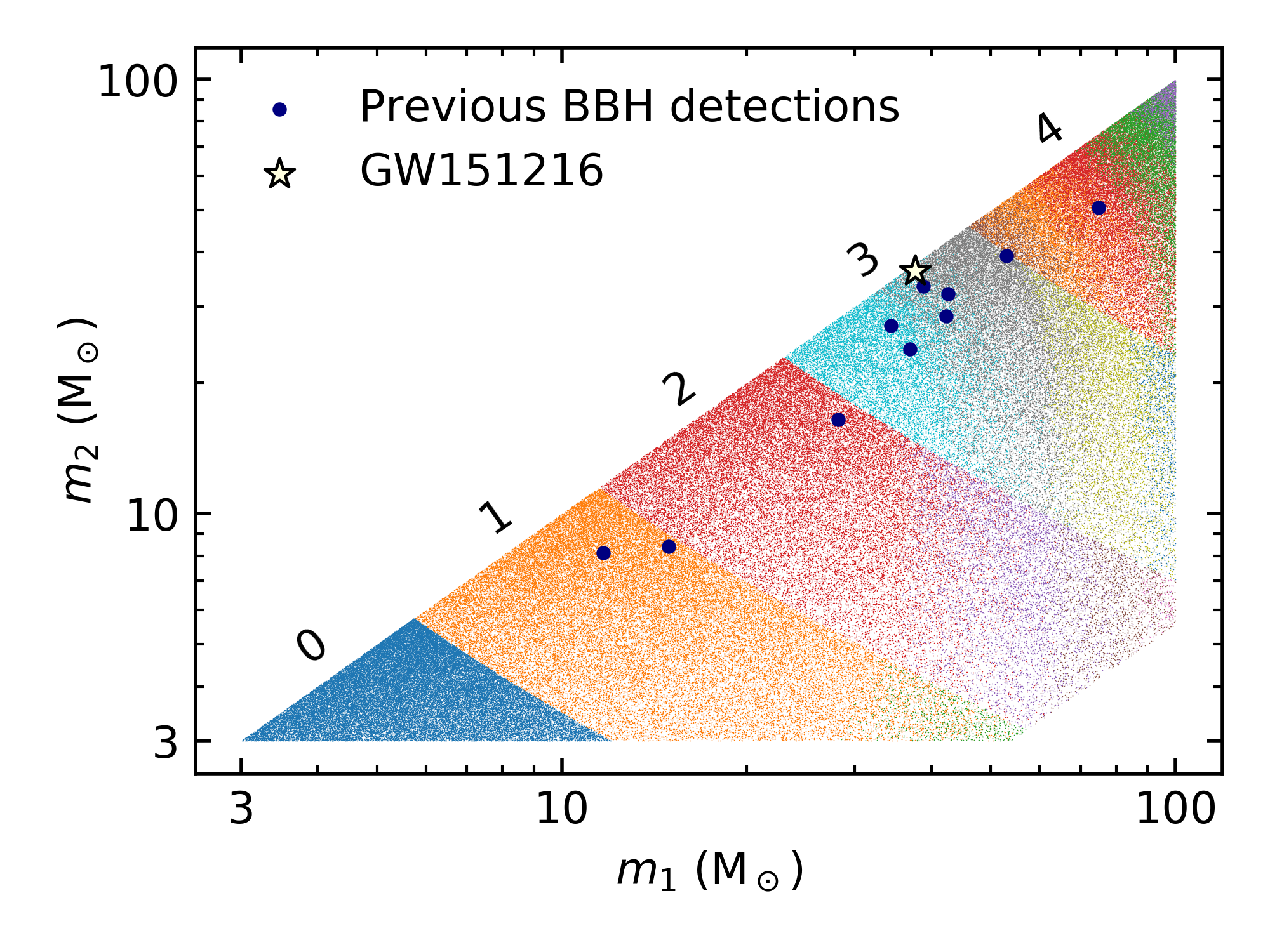}
    \caption{Division of the BBH parameter space into five template banks (\texttt{BBH 0-4}) by component masses. A separate search is conducted on each. The points represent the input waveforms used to construct the banks (not the templates themselves), and the colors encode the division of each bank into subbanks according to the shapes of the waveform amplitude. Approximate detector-frame masses are indicated for BBH detections reported to date (in O1 and O2) and for GW151216.
    }
    \label{fig:all_banks}
\end{figure}

As described in Ref.~\cite{templatebankpaper}, we construct five BBH template banks (\texttt{BBH 0-4}) that together span this target parameter space, and conduct a separate search within each of them. 
The banks are defined by regions in the plane of component masses, as shown in Fig.~\ref{fig:all_banks}. We place the bounds between adjacent banks at $\mathcal M = \{5, 10, 20, 40\}\, M_\odot$, where $\mathcal M = (m_1 m_2)^{3/5}/(m_1+m_2)^{1/5}$ is the chirp mass and $m_{1,2}$ are the individual masses.
We find several motivations for dividing the search. The low-mass banks have many more templates than the heavier banks, and thus they inherently have a larger look-elsewhere penalty. Dividing the search prevents this from strongly affecting the sensitivity of the high-mass searches: 
in this way, on astrophysical grounds we might expect roughly comparable numbers of signals in each bank, regardless of the largely different number of templates they have.
Moreover, this splitting enables us to discriminate between the different types of background events that each search is subject to.
The different duration of the signals in each bank will require us to use different thresholds when masking bad data segments (see Section~\ref{sec:badsegments}). The prevalence of non-Gaussian glitches will be different in each bank and thus the score we assign to events with the same signal-to-noise ratio (SNR) is different in each bank (see Section~\ref{sec:incoherent}).
Table~\ref{tab:banks} summarizes the template bank parameter ranges and sizes.

The template bank needs to be effectual, that is, to guarantee a sufficiently high match between a GW waveform and at least one template in the bank. We define the inner product between waveforms $h_i, h_j$
\begin{equation} \label{eq:ip}
    (h_i \mid h_j) \vcentcolon= 4 \int_0^\infty {\rm d} f\, \frac{\tilde h_i(f) \tilde h_j^\ast(f)}{S_n(f)},
\end{equation}
where $S_n(f)$ is the one-sided noise power spectral density (PSD) of the detector and a tilde indicates a Fourier transform into the frequency domain. It is used to define the match
\begin{equation}
    m_{ij} = \max_\tau \big| (h_i \mid h_j e^{i 2 \pi f \tau}) \big|;
\end{equation}
throughout this section we assume that all waveforms are normalized to $(h \mid h) = 1$. We assess the effectualness $E$ of each bank by computing the best match with \num{e4} random waveforms in its target parameter space. We apply the down-sampling and sinc-interpolation described in Section~\ref{sec:matchedfiltering} and the waveform optimization described in Section~\ref{sec:optimization} to the test waveforms, to properly simulate the search procedure. 
We report the effectualness of the banks in Table~\ref{tab:banks}.
When designing banks, we set the reference PSD to be the \texttt{aLIGO\_MID\_LOW} PSD \citep{lalsuite}, which is representative of O1.

\begin{table}
    \centering
        
    \begin{tabular}{l c *{2}{S[table-format=1.2]}S[table-format=6.]}
         \tabline
         \tabline
         Bank & $\mathcal M \,({M}_\odot)$ & {$E_0$} & {$E$} & {$N_{\rm templates}$} \\
         \tabline
         \texttt{BBH 0} & ${} < 5$ & 0.90 & 0.97 & 6465 \\
         \texttt{BBH 1} & $(5, 10)$ & 0.92 & 0.96 & 7919 \\
         \texttt{BBH 2} & $(10, 20)$ & 0.94 & 0.96 & 5855 \\
         \texttt{BBH 3} & $(20, 40)$ & 0.95 & 0.96 & 594 \\
         \texttt{BBH 4} & ${} > 40$ & 0.97 & 0.97 & 57 \\
         \tabline
         Total & & & & 20890
         \\
         \tabline
         \tabline
    \end{tabular}
    \caption{Summary of template bank parameters. $\mathcal{M}$ is the chirp mass range that the bank is designed to cover. $E_0$ and $E$ are the effectualnesses without and with refinement (Section~\ref{sec:optimization}) respectively, as quantified by the best match within the bank achieved by the top $99.9 \%$ of random astrophysical templates. $N_{\rm templates}$ is the total number of templates in each bank.}
    \label{tab:banks}
\end{table}

In order to correct the PSD drift at manageable computational cost, our search pipeline requires that the frequency domain templates, of the form
\begin{equation}
    \tilde h(f) = A(f)\, e^{i\psi(f)},
\end{equation}
share a common amplitude profile $\overline A(f)$ (see Section~\ref{sec:psd_drift}) and differ only in the phase $\psi(f)$. In order to avoid excessive loss of effectualness due to this approximation, we split each bank into several subbanks, each of which is assigned a different $\overline A(f)$ profile. We use the method of ``stochastic placement'' to determine as many subbanks as needed to guarantee that every waveform within the target parameter range has an amplitude match,
\begin{equation}
    \int{\rm d} f\, \frac{A(f) \overline A(f)}{S_n (f)} \geqslant 0.95,
\end{equation}
with at least one subbank. The resultant divisions into subbanks are color-coded in Fig.~\ref{fig:all_banks}.

The remaining task is to place templates in each subbank to efficiently capture the possible phase shapes $\psi(f)$. We achieve that with a geometric approach, where we use the mismatch between templates to define a mismatch distance, which quantifies the similarity between any two waveforms. We abandon the physical parameters as a description of the templates in favor of a new basis of coordinates $\bm{c}$, in which the mismatch distance induces an Euclidean metric. We then set up a regular grid in this space.
Our templates take the form
\begin{equation} \label{eq:calpha_decomposition}
    h(f; \bm{c}) = \overline A(f)\, \exp\Big[i \Big(\overline\psi(f) + \sum_\alpha c_\alpha\, \psi_\alpha (f) \Big)\Big],
\end{equation}
where $\overline{\psi}(f)$ is the average phase, and $\{\psi_\alpha(f)\}$ are phase basis functions which are orthonormalized such that the mismatch distance satisfies
\begin{equation} \label{eq:Euclidean}
\begin{split}
    d^2_{\bm{c}, \bm{c} + \bm{\delta c}} &\vcentcolon= 1 - m(h(\bm{c}), h(\bm{c} + \bm{\delta c})) \\
    &= \frac 12 \sum_{\alpha}\, \delta c_\alpha^2 + \mathcal O(\delta c^3).
\end{split}
\end{equation}
An input set of physical waveforms representing the target signals are used, first to define the subbanks and then to determine the appropriate phase basis functions. The input waveforms may be generated with any frequency-domain model; we use the \texttt{IMRPhenomD} approximant \citep{Khan2016}. The phase basis functions are found from a singular value decomposition of the input waveforms which identifies the minimal set of linear independent components that need to be kept. A small number of basis functions are enough to approximate all possible phases to sufficient accuracy. All banks require five linearly independent bases or fewer, with about half of them having only three or fewer. While the coefficient for the lowest order bases may vary over a range of several hundred units, the coefficients for the highest order bases vary within narrow ranges, sometimes by less than one unit.





\subsection{Loading and preprocessing the data}
\label{sec:preprocessing}


The strain data is provided by LIGO in sets of files of length 4096 seconds for each detector (H1 and L1 in O1). 
The natural choice is to split the analysis along the same lines, i.e., file by file.
We would like to preserve our sensitivity to events near the edges of files, and hence we pull in data from adjacent files if available. 
The length of data we pull in is set by the following considerations: (a) there should be no artifacts in the whitened strain at the edge of a file due to missing data at the right edge, (b) events that straddle files should be contained inside the padded and whitened data stream, and (c) relatively short segments of data ($< \SI{1024}{\second}$) near file edges, with a large adjoining segment ($> \SI{64}{\second}$) of missing data, are analyzed as part of the adjoining file instead of on their own.
Even after padding, the boundary of the (expanded) data stream will still have artifacts from the whitening filter. 
To treat this, we further append \SI{64}{\second} of zeros to the padded strain data on either side, that we will later inpaint using the method of Section~\ref{sec:holefilling}.

Additionally, we observe that long segments ($\gtrsim \SI{64}{\second}$) of bad data, as marked by LIGO's quality flags, can have a few unmarked extra seconds of bad data adjoining the marked segments (this can happen due to latency in the flagging system, for example). 
The procedure outlined in Section~\ref{sec:badsegments} is designed to catch such segments, as well as other kinds of misbehaved data. 
However, we only reach this stage after some initial signal processing and sufficiently bad data segments might pollute good data segments through each step of the analysis.
Therefore, we trim an additional \SI{2}{\second} of data when these segments occur at the right edges of files.

The next step after loading the data is to estimate its PSD. 
We use Welch's method~\cite{welch1967use}, in which several overlapping chunks of data are windowed and their periodograms are averaged (we use the implementation in \texttt{scipy.signal} with a Hann window). 
We make our PSD estimation robust to bad data by (a) disregarding chunks that overlap with segments that were marked by LIGO's quality flags, and (b) averaging using the median instead of the mean (see Appendix B of Ref.~\cite{Allen2012}). 

An important choice to make is the length of the individual chunks whose periodograms enter the averages (`chunksize' in what follows). 
In pure Gaussian random noise, the choice of chunksize is governed by the following (conflicting) considerations: (a) controlling the statistical uncertainty in the averages, which depends on the number of independent samples within a file, and (b) mitigating the loss in matched-filter sensitivity around under-resolved spectral lines.
As we discuss in Section~\ref{sec:psd_drift}, the advanced LIGO data is typically not described by purely Gaussian random noise (even in the absence of ``bad" segments with excess power) due to systematic drifts in the PSD within a file. 
We find that using \SI{64}{\second} chunks to measure the PSD yields an acceptable compromise between the above effects. 
This choice also affects the minimum length of the files that we choose to analyze: the first consideration above (the measurement noise in the PSD) implies that we take a $4 \%$ loss in sensitivity for files that are shorter than 16 times the chunksize. 
If a file is shorter than this limit (not including the segments marked by LIGO's quality flags), we try to analyze it using a chunksize of \SI{16}{\second} instead, while enforcing the same minimum number of chunks. 

We restrict ourselves to analyzing frequencies $f < \SI{512}{\hertz}$ by down-sampling the data to $\SI{1024}{\hertz}$. 
This is safe to do since all compact binary merger signals accumulate more than $\simeq 99 \%$ of their matched filtering SNR below $\SI{512}{\hertz}$ at the O1 detector sensitivity\liang{ (BNS mergers accumulate more than $\simeq 99\%$ of the SNR)}, and since we already budget for $\gtrsim 1\%$ losses in the template bank. 
This choice reduces the sizes of the template banks and saves us computational time during triggering, at the expense of a negligible loss in sensitivity.
We also apply a high-pass filter to the data (implemented as a fourth-order Butterworth filter with $f_{\rm min} = \SI{15}{\hertz}$, applied from the left and the right to preserve phases). 
This removes low-frequency artifacts in the data (that could later trigger our flagging procedure in Section~\ref{sec:badsegments}), and is safe to do since we only use frequencies $f > \SI{20}{\hertz}$ in building the template bank.

Finally, we construct the whitening filter from the estimated PSD, and use it to whiten the data. 
The whitening filter typically has most of its power at small lags, but exhibits a long tail at large lags due to spectral lines in the data. 
Our procedure for inpainting bad data segments (described in Section~\ref{sec:holefilling}) requires that the whitening filter has finite support, hence we zero the filter at large lags (while ensuring that we retain $\gtrsim 99.9 \%$ of its weight, typically the filter is left with an impulse response length of $\simeq \SI{16}{\second}$).
Zeroing the whitening filter in the time domain corresponds to convolution with a sinc function in the frequency domain, which fills in the lines; thus, the filter does not reject spectral lines completely.
Hence, we take care that our flagging procedure does not trigger on spectral lines in the data.


\subsection{Identifying bad data segments}
\label{sec:badsegments}
Advanced LIGO data contains intermittent loud disturbances that are not marked by the provided data quality flags. 
We need to flag and remove these segments to prevent them from polluting our search, while taking care to preserve astrophysical signals of interest. 
This is the fourth analysis of the data, and hence we assume that any new signals we find will have an integrated matched filter SNR $\rho < 30$ in a single detector. 
This assumption allows us to bound the influence of a true signal on our procedure.

We devise several complementary tests to flag bad data segments. We design our tests to satisfy the following conditions:
\begin{enumerate}
\item The test statistics have analytically known distributions for Gaussian random noise.
\item The thresholds are set to values of the test statistics achieved by waveforms with single-detector $\rho = 30$ in noiseless data.
Signals at this SNR have a probability of $\simeq 0.5$ of triggering a single test in the presence of Gaussian random noise. We found empirically that signals satisfying $\rho \leq 20$ are almost always retained.
\item If the above thresholds are too low, they are adjusted so that a single test is triggered at most once per five files due to Gaussian random noise alone. This is important for template banks with long waveforms.
\item The tests are safeguarded from being triggered by PSD drifts over long timescales ($t \gtrsim \SI{10}{\second}$), which can manifest as excess power over shorter timescales.
\end{enumerate}
These conditions ensure that we are sensitive to gravitational waves while not over-flagging the data.
It is important that the tests be done at the single-detector level in order to avoid biasing the calculation of the background using time slides.

Our tests trigger on the following anomalies: (a) outliers in the whitened data-stream, (b) sine-Gaussian transients in particular bands, (c) excess power localized to particular bands and timescales, and (d) excess power (summed over frequencies) on particular timescales. 
We picked timescales and frequency bands for the tests based on inspecting the spectrograms of the bad segments; Table~\ref{tab:data_quality_checks} details the choices.

\begin{table}
    \centering
    \caption{Summary of tests for identifying bad data segments. For each test, we show the frequency band and timescale of the disturbance that it is sensitive to, and the length of the data we excise around the disturbance.}
    \setlength\tabcolsep{1pt}
    \begin{tabular}{l c *{2}{S[table-format=1.3]}}
        \tabline
        \tabline
         Test type & \head{2cm}{Frequency band} & \head{2cm}{Excess duration (s)} & \head{2cm}{Hole duration (s)} \\
         \tabline
         Whitened outlier
          & $[20, 512]$ & {$10^{-3}$} & 0.6 \\
         \hline
          & $[20, 512]$ & 0.2 & 0.2\\
          & $[20, 512]$ & 1 & 1 \\
          & $[55, 65]$ & 1 & 1 \\
          & $[70, 80]$ & 1 & 1 \\
          & $[40, 60]$ & 1 & 1 \\
         Excess power
          & $[40, 60]$ & 0.5 & 0.5 \\
          & $[20, 50]$ & 1 & 1 \\
          & $[100, 180]$ & 1 & 1 \\
          & $[25, 70]$ & 0.1 & 0.1 \\
          & $[20, 180]$ & 0.05 & 0.05 \\
          & $[60, 180]$ & 0.025 & 0.025 \\
          & $[25, 70]$ & 0.2 & 1 \\
         \hline
          & $[55, 65]$ & {-} & 0.1 \\ 
          & $[20, 60]$ & {-} & 0.1 \\
          & $[100, 140]$ & {-} & 0.1 \\
         Sine-Gaussian\footnote{Sine-Gaussian transients saturate the uncertainty principle, and hence their duration is fixed given their bandwidth.}
          & $[50, 150]$ & {-} & 0.1 \\
          & $[70, 110]$ & {-} & 0.1 \\
          & $[50, 90]$ & {-} & 0.1 \\
          & $[125, 175]$ & {-} & 0.1 \\
          & $[75, 125]$ & {-} & 0.1 \\
         \tabline
         \tabline
    \end{tabular}
    \label{tab:data_quality_checks}
\end{table}

The data has spectral lines at which the PSD is several orders of magnitude higher than in the continuum. 
The power in these lines often significantly varies in a non-Gaussian manner within a single file. 
The lines do not contribute to the matched-filtering overlap, since the PSD is effectively infinite at their frequencies. 
Hence it is preferable that varying lines do not trigger our tests. 

We detect sine-Gaussian artifacts in a given band by matched-filtering with a complex waveform that saturates the time-frequency uncertainty principle and contains most of its power in the band. 
We apply notch filters to the sine-Gaussian template to remove any overlap with spectral lines. 
We flag any outliers in the matched-filtering results above a threshold defined to satisfy the aforementioned conditions (see second paragraph of \refsec{badsegments}), which is a procedure safe to any relevant events)


We detect excess power using a spectrogram (computed using the \texttt{spectrogram} function in \texttt{scipy.signal} with its default Tukey window). 
We sum the power in the frequency ranges of interest, disregarding frequency bins that overlap with varying lines. 
For Gaussian random noise, this sum has a chi-squared distribution. 
This is not achieved in practice unless correcting for the effects of PSD changes.
We make the excess power statistic robust to the drifting of the PSD by comparing the instantaneous excess power with with a local moving-average power baseline.

The simplest test is to look for outliers in the whitened strain, since individual samples should be independent and normally distributed with unit variance. We flag segments of whitened data, with a safety margin in time, around outliers above a chosen threshold.

Whenever one or more of these tests fire, we excise the offending segments (which we refer to as ``holes'') and inpaint the raw data within as described in Section~\ref{sec:holefilling}. 
In practice, we observe that the outlier test often does not catch all of the ``bad" data, in which case the inpainted and whitened data contain further outliers. 
Hence, we iterate over the ``identify bad segments, inpaint, whiten" cycle multiple ($< 7$) times, increasing the safety margin in time by successively larger multiples of 0.1 s, until the process converges. 



We treat any part of the data that was marked with any of the LIGO quality flags as if it contained large disturbances.
After all the data quality tests done in this section, we are left with roughly 46 days of coincident on-time between the detectors, with slight changes from bank to bank, as all the test thresholds are waveform dependent.

\subsection{Inpainting bad data segments}
\label{sec:holefilling}

The matched-filtering score for a template $h$ with data $d$ with a noise covariance matrix $C$ is:
\begin{align}
    Z =h^{\dagger}\,C^{-1}\,d = 4\,\sum_{f}{\frac{h^*(f)\,d(f)}{S_n(f)}}, \label{eq:matchedfiltereq}
\end{align}
where $f$ denotes the frequencies, and in the last equality we assumed that the noise is diagonal in Fourier space. 
The tests described in Section~\ref{sec:badsegments} flag bad data segments that we would like to mask.
The operator $C^{-1}$ (the ``blueing filter") is not diagonal in the time domain; when viewed as a linear filter operating on the data, its impulse response length (typically $\simeq \SI{32}{\second}$) is set by the PSD spectral lines and the chunktime used to estimate the PSD. 
Thus the scores evaluated using Eq.~\eqref{eq:matchedfiltereq} can be significantly affected even tens of seconds away from a masked segment.

To deal with this problem, if we consider a fraction of the data of length $N_d$ in which we have masked $N_{h}$ samples, we filter the data with a filter $F$ and define a new score by:
\begin{equation}
    \tilde Z=h^{\dagger}\,C^{-1}\,F\,d. 
\end{equation}
The filter $F$ is given by 
\begin{equation}
    F = 1 - W\, M^{-1}\, W^{T}\, C^{-1},
\end{equation}
where the matrix $W$ has one column of length $N_d$ for every sample that is masked with all the entries zero except for a one at the position of the masked sample and $M$ is the $N_h\times N_h$ matrix $M=W^T\,C^{-1}\,W$. The computationally expensive part of this filtering procedure is to invert the matrix $M$.

The filter $F$ is such that the score $\tilde Z$ is independent of the value of the template waveform $h$ inside masked segments. 
That is to say, $F$ can be obtained by demanding that $C^{-1} F d$ is identically zero inside the masked regions. 
$F$ is a projection operator ($F^2=F$) that commutes with $C^{-1}$, i.e., $C^{-1} F=F^T\,C^{-1}$, and depends only on the mask and the covariance matrix $C$. 
In particular, it is independent of the waveform $h$, and thus can be computed once and for all before performing matched filtering. 
Note also that for computing $F$, it is not important that $C^{-1}$ be the exact noise covariance; it just needs to be consistently used to define the scores in the section of data. 

We can also derive $F$ as the solution of several related linear algebra problems. We can model the presence of the mask as if the data had an additional source of noise inside the masked region, and take the limit of zero additional noise outside the holes and infinite additional noise inside. The filtered data $\tilde d=F\,d$ equals the original data outside the masked segments, and the best linear prediction for the data inside the hole based only on the data outside (Wiener filter). It can also be thought of as the $\tilde d$ that minimizes
\begin{equation}
    \chi^2=\frac{1}{2}\, \tilde d^\dagger\,C^{-1}\,\tilde d
\end{equation}
subject to the constraint that $\tilde d$ equals the original data outside the mask, but can take any value inside. The computation of $F$ is explained in detail in Ref.~\cite{psddriftpaper}.

\begin{figure}
    \centering
    \includegraphics[width=\linewidth]{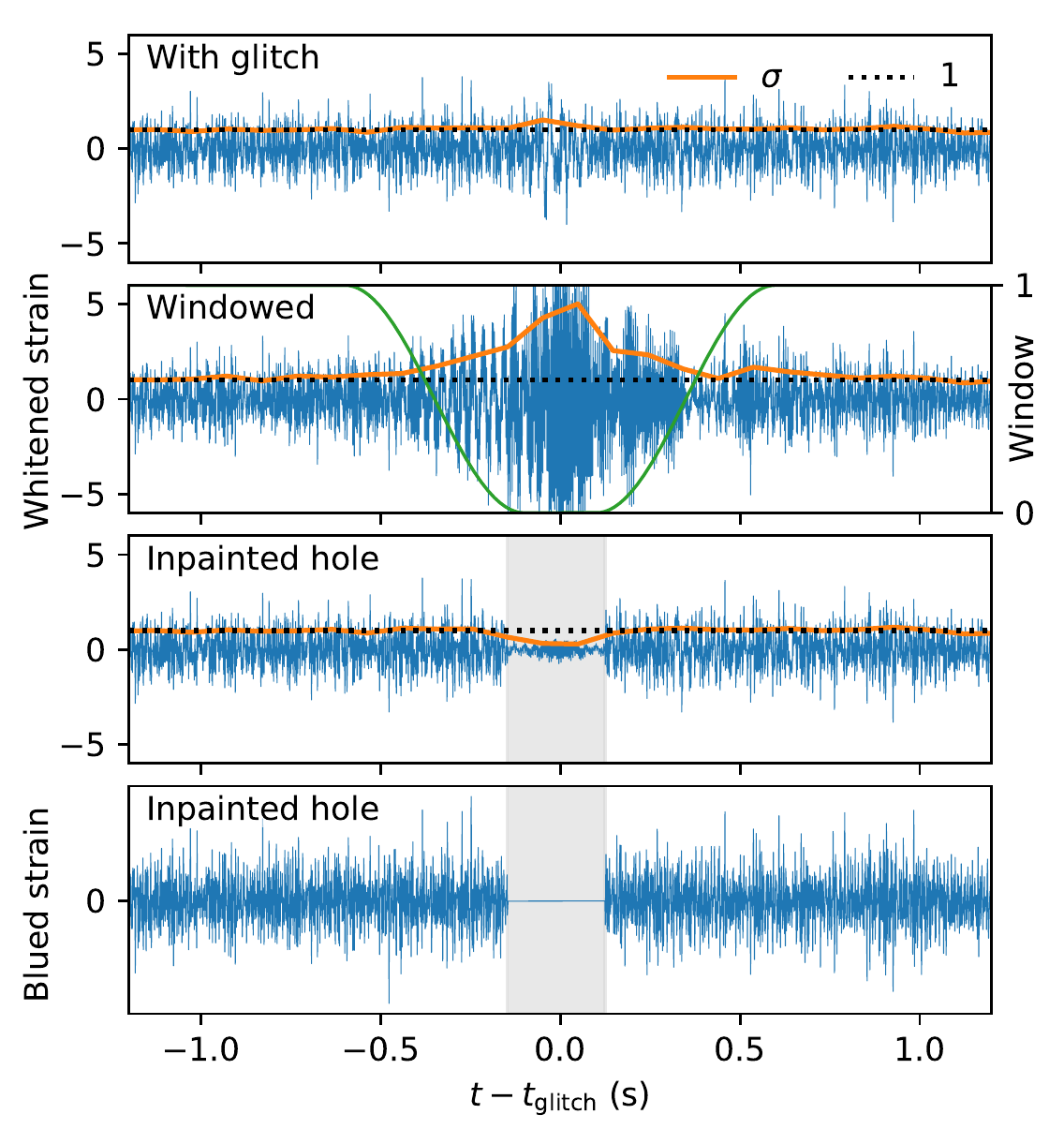}
    \caption{Effect of masking and inpainting glitches. \textit{Top panel:} A segment of whitened strain data (in units of the noise standard deviation) that has an identified glitch. The orange line is the standard deviation $\sigma$ over a running window of 100 samples, and is typically close to unity as expected for whitened data. \textit{Second panel:} Gating the glitch with an inverse Tukey window (green) and then whitening generates artifacts in the whitened data, even outside the window. For example, $\sigma$ remains above 1.1 for approximately \SI{2}{\second} to each side of the glitch. \textit{ Third panel:} The inpainted whitened data has unit variance outside the hole (shaded). \textit{Bottom panel:} After inpainting, the ``blued" strain is identically zero inside the hole, so overlaps with templates do not depend on what is inside the hole.}
    \label{fig:hole}
\end{figure}

Figure \ref{fig:hole} shows an example of a small section of the data containing a ``glitch" artifact. We show the difference between `gating' the bad data by applying a window function to it, and creating a hole and inpainting it with the algorithm we described. We can see that gating substantially changes the standard deviation of the samples in the hole and the few seconds surrounding it, which can potentially create spurious triggers, and can damage any real signals that happen to be in the data at the same time. In our method, the ``blued" data is set to be identically zero inside the hole. 
\subsection{Matched filtering}
\label{sec:matchedfiltering}

Given the whitened, hole-filled data, we compute the overlaps with all templates in the template bank, and register the times and templates when the ${\rm SNR}^2$ is above a triggering threshold. 
The choice of the threshold was driven by the requirement to produce a manageable number of triggers per file, and was generally in the range $20 < {\rm SNR}^2_{\rm thresh} < 25$ for the various banks and subbanks.

In order for the statistics of the overlaps to have a standard complex normal distribution,
we need to apply two corrections: one is for the PSD drift, and one for the existence of holes (masked data segments).
As we show in Ref.~\cite{psddriftpaper}, the PSD correction depends only on the amplitude of the waveform, and hence we pre-compute it for each representative $A(f)$. 
The hole correction is waveform dependent: we evaluate it under the stationary phase approximation, which assumes that there are many waveform cycles inside the hole, and accounts for the change in the variance due to the missing cycles in the hole. 
This approximation works only for long waveforms, and hence we use overlaps in the vicinity of holes only for waveforms that are longer than \SI{10}{\second}.
We also ignore overlaps where more than half of the variance (and hence ${\rm SNR}^2$) is inside holes as these are anyway a negligible part of the volume (and are also non-declarable even if they contain a genuine candidate).

In order to compute the overlaps and hole variance corrections efficiently, we first notice that the waveform is shorter than a typical data segment, so we can use the overlap--save method in order to reduce the FFT sizes.
Because the maximum frequency of the whitened data is taken to be $\SI{512}{\hertz}$, all information about matching the template to the data is in the complex overlaps we compute.
Looking at single overlaps and comparing to the triggering threshold is not sufficient since the SNR could be reduced by as much as 10\% due to sub-sample shifts in the GW arrival time (we down-sampled the data to $\SI{1024}{\hertz}$). We recover this sensitivity by first setting a lower SNR bar, and sinc-interpolating the overlaps (by a factor of 4) within each contiguous segment above this lower bar, before checking for overlaps above the (higher) triggering threshold.

\subsection{Applying corrections due to the varying Power Spectral Density of the Noise} \label{sec:psd_drift}

\begin{figure}
    \centering
    \includegraphics[width=\linewidth]{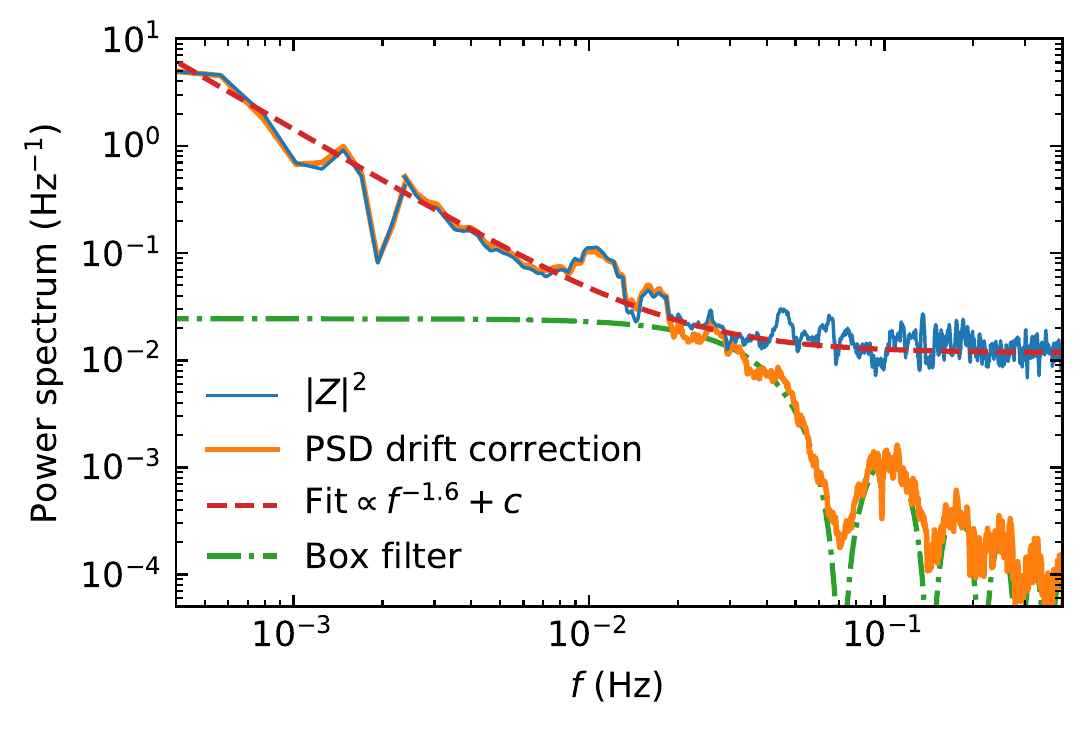}
    \caption{It is necessary to track the drifting PSD on time scales of seconds. In blue we show the power spectrum of the square of the absolute value of the overlaps with a template in the \texttt{BBH 0} bank for a repesentative set of files. It reaches the level of Gaussian fluctuations only close to $\sim \SI{0.1}{\hertz}$, and has a red-noise power spectrum fit by a power-law (red dashed curve). The orange curve shows the PSD drift correction we apply to the data, which correctly traces the actual fluctuations in the standard deviation of the overlaps up to the Gaussian floor.}
    \label{fig:ps_psd_drift}
\end{figure}

The power spectral density of the LIGO detectors can slightly vary with time. These changes may be hard to track and would inevitably result in PSD mis-estimation. As Ref.~\cite{psddriftpaper} shows, if we mis-estimate the PSD by a factor $(1+\epsilon(f))$, the information loss in matched filtering scales as $\mathcal{O}(\epsilon^2)$, but the overlap's standard deviation differs by $\mathcal{O}(\epsilon)$.
This means that $\mathcal{O}(100)$ segments of data are required in order to measure the PSD well enough to aim for discarding less than $1\%$ sensitivity.
In order to resolve the lines well enough to aim for the same loss, tens of seconds of data are required. Therefore, an order of a thousand seconds are needed for estimating the PSD.
We choose to measure the PSD using the Welch method, in which the signal is cut into overlapping segments, and the PSD power at frequency $f$ is the (scaled) median of all the power estimates at this frequency from all the segments. 
It turns out, though, that the PSD varies on time-scales as short as $\sim\SI{10}{\second}$, as seen in Figure~\ref{fig:ps_psd_drift}.

While at first sight it may seem impossible to both capture the width of the lines and track the fast variation in the PSD, we accomplish it by correcting the first order effect of PSD mis-estimation on time-scales that are as short as the PSD changes, to precision of $\sim 1\%$.

\begin{figure}
    \centering
    \includegraphics[width=\linewidth]{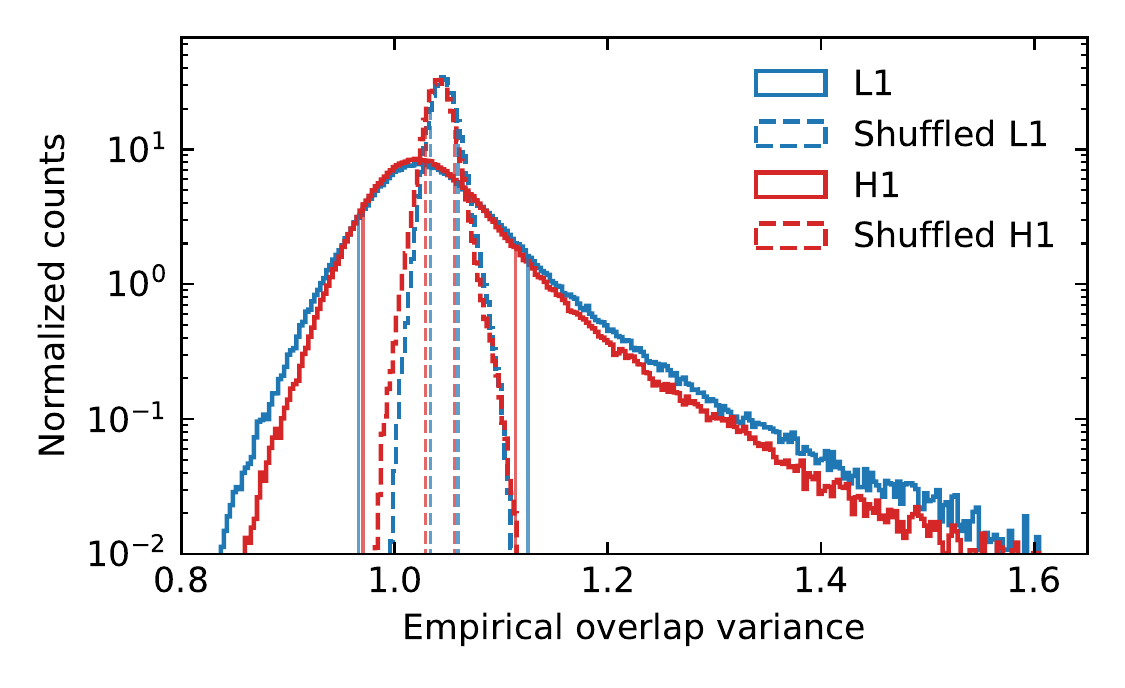}
    \caption{Estimated changes to the variance of the overlap measurements, measured over periods of $\approx\SI{16}{\second}$ defined to guarantee a 2\% precision. Measurement errors are shown by shuffling the overlaps in time and calculating the local averages. Vertical lines are one standard deviation away from the mean for each distribution. It is evident that the variance changes we are tracking are not random measurement fluctuations and can lead to severe changes in the significance assessment of a particular event.}
    \label{fig:psd_drift_dist}
\end{figure}

This correction is basically a local estimate of the standard deviation of the overlaps, and is derived (along with some other nice properties it has) in Ref.~\cite{psddriftpaper}.
In Figure~\ref{fig:psd_drift_dist}, we present a histogram of the distribution of the local variance estimates. Notice the large deviations from unity in both directions. We note that the tail reaches values as high as $1.5$; at such high values, there are visible disturbances in the spectrogram, sometimes referred to as glitches.
However, at values in the range $[0.85,1.2]$, the data mostly behaves in a regular fashion, and there is no apparent sign something bad is going on in the spectrogram of the data. These changes can cause substantial loss of sensitivity in binary coalescence analyses that neglect this effect\footnote{After this manuscript was made public, we were informed that fluctuations in the SNR integral (due to short-timescale variations in the PSD) at comparable levels were previously noted, but the mitigation steps were not incorporated into the search pipelines used in the catalog paper (Thomas Dent, private communication).}. 

\begin{figure}
    \centering
    \includegraphics[width=\linewidth]{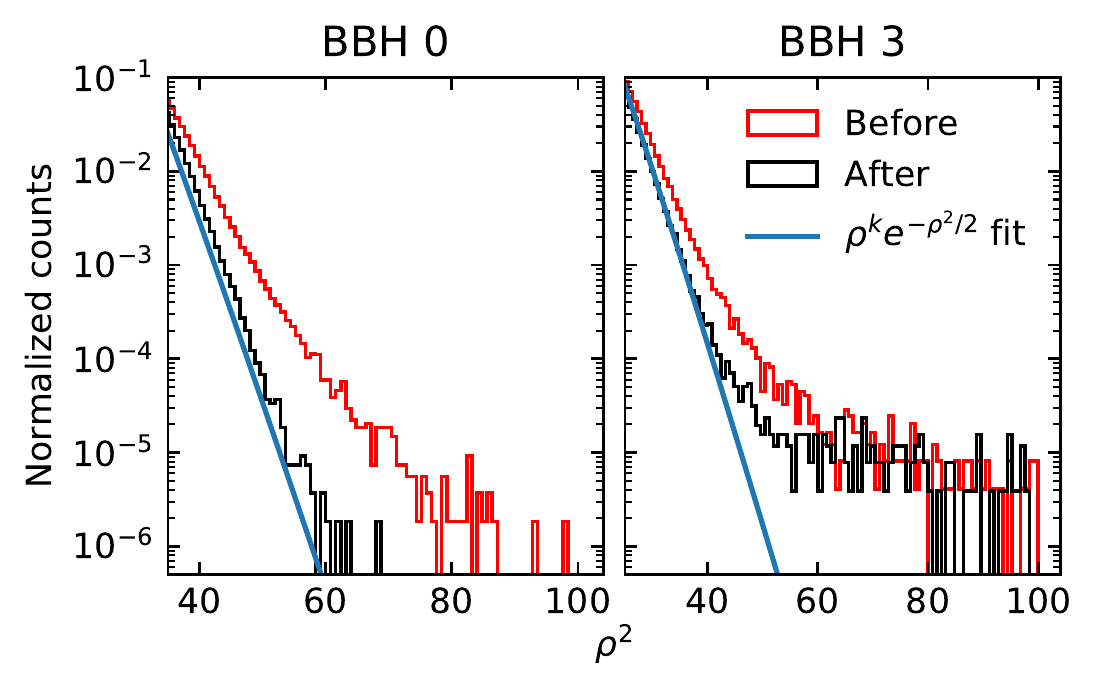}
    \caption{Effect of the PSD drift correction on the trigger distribution. Trigger distributions of binary black hole merger waveforms in bank \texttt{BBH 0} ($\mathcal{M} \in [2.6,5]\,M_{\odot}$) and a subbank from \texttt{BBH 3} ($\mathcal{M} \in [20,40]\,M_{\odot}$), in the Hanford detector, before applying any vetoes.
    }
    \label{fig:trigger_distribution_after_psd_drift_corr}
\end{figure}

To illustrate why correcting for these variance estimates is crucial for determining the exact significance of a candidate event, we point out that the most economic way of creating a (spurious) $\rho=8$ event is to wait for a lucky time where the PSD mis-estimation is large (say, 1.2), and then create a (genuine) $\rho=7.3$ fluctuation. In Figure~\ref{fig:trigger_distribution_after_psd_drift_corr}, we see the tail of the trigger distribution is substantially inflated if the PSD drift is not corrected.

\subsection{Coincidence Analysis of the two detectors}
\label{sec:coincidence}

After all single detector triggers above a critical $\rho^2$ are collected, we need to find pairs of triggers that share the same template, and have a time-lag difference that is less than $10\,{\rm ms}$.
In order to generate background coincident triggers, we also need to collect trigger pairs with all other considered time slides (we choose integer jumps of $0.1\,{\rm s}$ in the range $[-1000\,{\rm s},1000\,{\rm s}]$).
We collect the background events and the physical events by the following process:
First, we define that a real trigger has $\rho^2 > 0.9\,\rho^2_{\max} - 5$ where $\rho_{\max}$ is the maximum trigger in the segment of $\SI{0.01}{\second}$.
The reason for this choice is that triggers that are too close to a major erratic event are not declarable and that if there is a glitch that slipped through our net, we do not want a large amount of accompanying triggers to coincide with random fluctuations in the other detector. This massively reduces the load of the subsequent stages.

We then take each remaining trigger, and insert it into a dictionary according to the template key. This would allow us to immediately find all the times at which this template triggered.
Using queries to the dictionary, we find all the pairs of triggers that belong to either the background or the foreground group, and pass the threshold $\rho_{\rm collect}$. This threshold depends on the bank via computing the Gaussian noise threshold for obtaining one significant event per O1, and then multiplied by the bank effectualness, to guarantee that every trigger that can acquire the one-per-O1 significance after optimization is included.

We now view the H1 component of all pairs of triggers and group them to groups of 0.1s. We use the less stringent version of the veto to vet the trigger with the highest SNR in each group, and upon failure discard the entire group (the logic here is that similar triggers are all passing or failing the veto together).
We do the same for the L1 component of all remaining trigger pairs.

We then optimize every trigger by computing the overlaps with the data of every template in the sub-grid $\bm{c}$ values (see Sec.~\ref{sec:optimization}). We further sinc interpolate with a long support to obtain further time resolution for the overlaps. We then choose the sub-grid template that maximizes the quadrature sum of the single detector SNRs.
This trigger pair is now vetoed with the stringent veto.
If a trigger pair passes all these, it is registered.

 \subsection{Refining triggers on a finer template grid}
 \label{sec:optimization}
 
 The template-bank is organized as a regular grid, which facilitates refinement in places of interest. This enables us to squeeze more sensitivity and imitate the strategy of a continuous template bank, which is more objective than an arbitrarily chosen grid.
 The effectualness achieved by the top 99.9\% of injections with the template banks used for the search varies between 0.9 and 0.96. 
 Refining the grid by a factor of two in each dimension would bring it to $>0.96$ in all cases, but would also substantially increase the number of waveforms in the bank (which in turn increases the computational complexity and memory requirements of our search). 
 We therefore take the approach of refining every candidate and background trigger pair. 
 Since we know the maximum amount of SNR increase that is possible for a real event, we refine all candidates that have a score that is high enough to have a chance of reaching a FAR of 1/O1 after refinement. 
 We greatly speed up the candidate refinement by calculating the likelihood using the relative binning method~\cite{Zackay2018} (using the original grid-point trigger as the reference waveform).
 Table~\ref{tab:banks} reports the improvement in effectualness achieved by this procedure for our banks.

\subsection{Vetoing triggers}
\label{sec:vetoes}

The matched-filtering score is the optimal statistic for detecting signals buried in Gaussian random noise. 
As emphasized in the previous sections, the LIGO strain data is not well-described by purely Gaussian random noise, and hence, the matched-filtering score may be triggered (i.e., pushed above the Gaussian-noise significance threshold) by either transient or prolonged disturbances in the detector. 
Our pipeline attempts to reject these candidates by identifying bad segments at the preprocessing level (Section~\ref{sec:badsegments}), or downweighting the scores by their large (empirically measured) variance (Section~\ref{sec:psd_drift}). 
However, this is not enough to bring us down to the Gaussian detection limit, especially for the heavier black hole banks.
Thus, we need additional vetoes at the final stage to reject glitches. 
We use vetoes that are based on either the quality of the neighboring data, as well as that of the signal.

Our most selective vetoes are based on signal quality, and check that the matched-filtering SNR builds up the right way with frequency. We perform the following tests:
\begin{enumerate}
    \item 
    We subtract the best-fit waveforms from the data and repeat the excess power tests of Section~\ref{sec:badsegments}, but with lower thresholds computed using waveforms with $\rho = 3$ (and bounded to fire once per 10 files due to Gaussian noise).
    Moreover, when we see excess power in a particular band and at a particular time, we only reject candidates with power at the same time in their best-fit waveforms (in order to avoid vetoing candidates due to unrelated excess power). 
    \item We split the best-fit waveform into disjoint chunks, and check for consistency between their individual matched-filtering scores. 
    This test is similar in philosophy to the chi-squared veto described in Ref.~\cite{2005PhRvD..71f2001A}, but improves upon it by accounting for the mis-estimation of the PSD (which is an inevitable consequence of PSD drift) and by projecting out the effects of small mismatch with the template bank grid. 
    \item We empirically find triggers that systematically miss the low-frequency parts of the waveforms, or have large scores at intermediate frequencies. The check described above is agnostic to the way the matched-filtering scores in various chunks disagree, and hence is not the most selective test for these triggers. We reject these triggers by using ``split-tests" that optimally contrast scores within two sets of chunks.
\end{enumerate}
The final two tests are the most selective vetoes, and hence their thresholds must be set with care. 
Our method for constructing template banks enables us to set these thresholds in a rigorous and statistically well-defined manner to ensure a given worst-case false-positive probability, which, accounting for the inefficiency in the bank, is achieved with adversarial template mismatches. 
Hence we set the worst-case false-positive probability of $10^{-2}$ for each of these tests.
The details of the tests, and the methods to set thresholds, are described in Ref.~\cite{vetopaper}. We note that all hardware injections that triggered passed the single-detector signal-based veto.

The data-quality vetoes are relatively simple in nature, and motivated by segments with excess power (as observed in spectrograms) that slip through the combination of the flagging procedure (of Sec.~\ref{sec:badsegments}) and PSD drift correction (of Sec.~\ref{sec:psd_drift}). The tests are as follows:
\begin{enumerate}
    \item Sometimes, our flagging procedure only partially marks the bad segments, in which case short templates (such as those of the heavier black hole banks) can trigger on the adjoining unflagged data. 
    This is mitigated by our choice, described in Section~\ref{sec:matchedfiltering}, to discard candidates with short waveforms in the vicinity of holes in our data (in practice, we reject waveforms $< \SI{10}{\second}$ long within \SI{1}{\second} of a hole). 
    \item There are rare bad segments on timescales of $\simeq 5-\SI{10}{\second}$, which is too long for our flagging procedure but too short for the PSD drift correction. 
    We flag segments of duration \SI{25}{\second} with a statistically significant number of loud triggers ($\rho^2 \gtrsim 30$) that are local maxima within subintervals of \SI{0.1}{\second}. 
    We set a generous threshold that should be reached at most once per run (approximately accounting for correlations between templates) within Gaussian noise, and is robust to astrophysical events (due to the maximization over time). 
    \item Finally, we account for rare cases with significant PSD drifts on finer timescales than the ones used while triggering (described in Section~\ref{sec:psd_drift} and Ref.~\cite{psddriftpaper}). 
    When this PSD drift is statistically significant, we veto coincidence candidates (both at zero-lag and in timeslides) whose combined incoherent scores, after accounting for the finer PSD drift correction, are brought down below our collection threshold.
\end{enumerate}

\begin{figure}
    \centering
    \includegraphics[width=\linewidth]{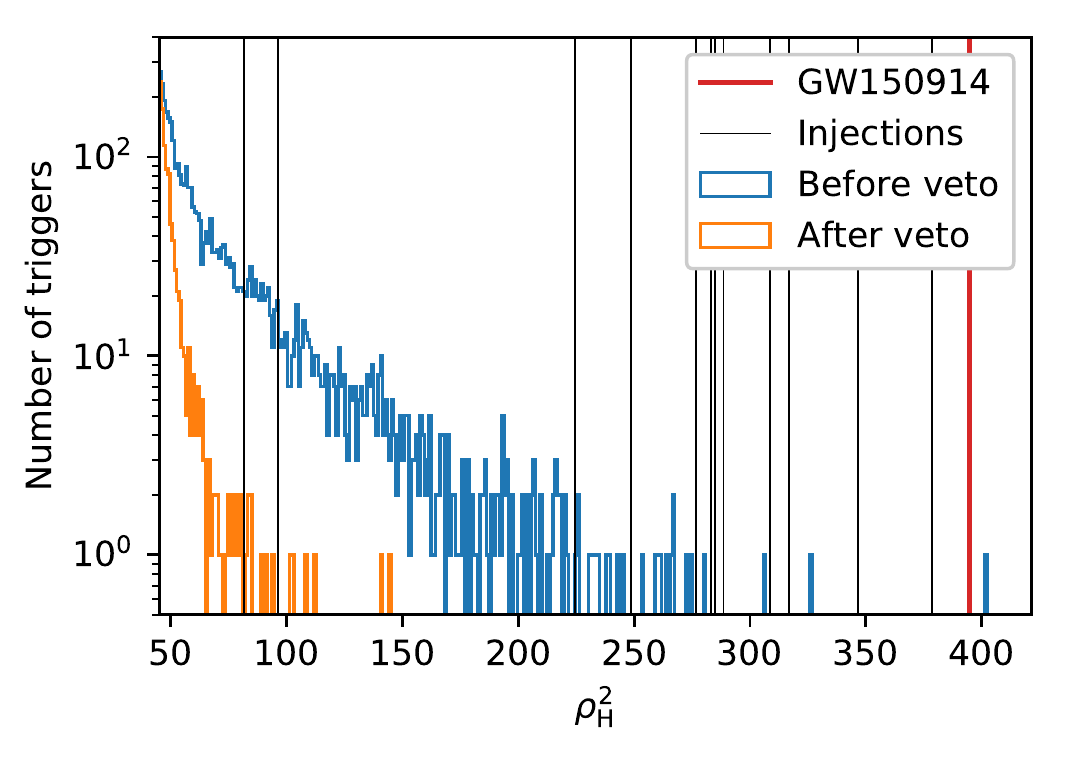}
    \caption{The impact of signal and data quality vetoes on the distribution of Hanford detector triggers in the \texttt{BBH 3} bank.
    GW151216 is deep in the Gaussian part of the distribution with $\rho_{\rm H}^2 = 39.4$, and is not shown in this plot.}
    \label{fig:veto_plot}
\end{figure}

Figure \ref{fig:veto_plot} shows the cumulative effect of our vetoes on the score distribution of the triggers in the \texttt{BBH 3} bank, which contains short waveforms of heavy binary black hole mergers.
Also shown are the hardware injections present in the data stream and GW150914 which belongs to this bank's chirp mass range.
We note that the veto retained every hardware injection in this chirp mass domain that passed the flagging procedure of Section \ref{sec:badsegments}.
It is interesting to note that GW150914 does not stand out from the\textit{ single detector} trigger distribution before the application of the veto, and is clearly detected even without resorting to coincidence after it.

\subsection{Incoherent Ranking}
\label{sec:incoherent}

\begin{figure}
    \centering
    \includegraphics[width=\linewidth]{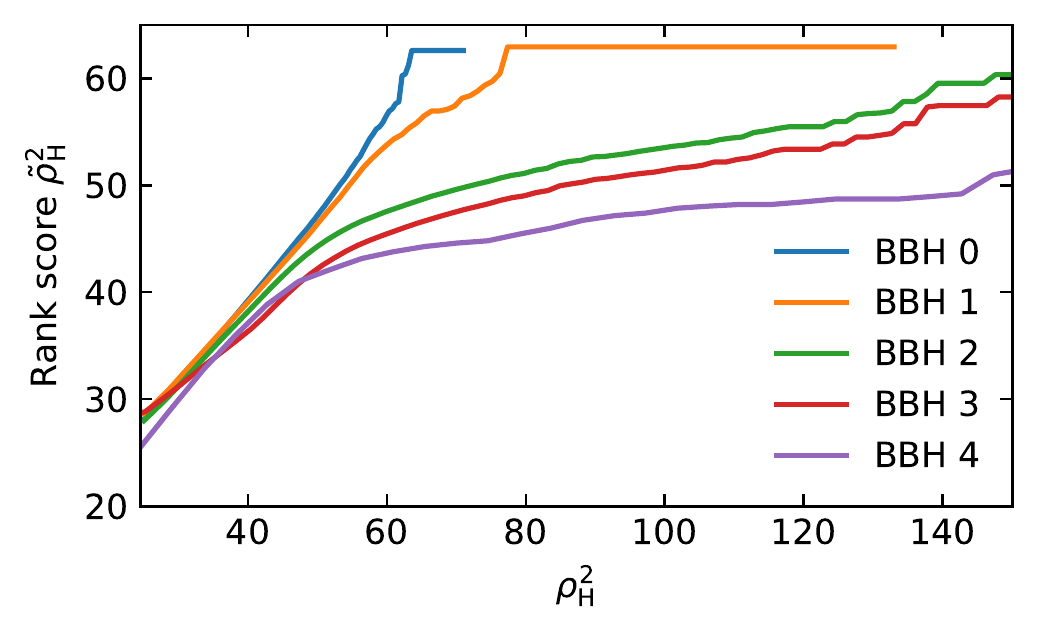}
    \caption{Relation between our new rank-based score $\tilde \rho$ and the SNR $\rho$, for the Hanford detector. The initial linear dependence reflects the Gaussian part of the trigger distribution, the curve saturates due to the non-Gaussian glitch tail. This effect is more prominent in the higher-mass banks, which are more sensitive to glitches.}
    \label{fig:rank_vs_snr}
\end{figure}

\begin{figure*}
    \centering
    \includegraphics[width=\linewidth]{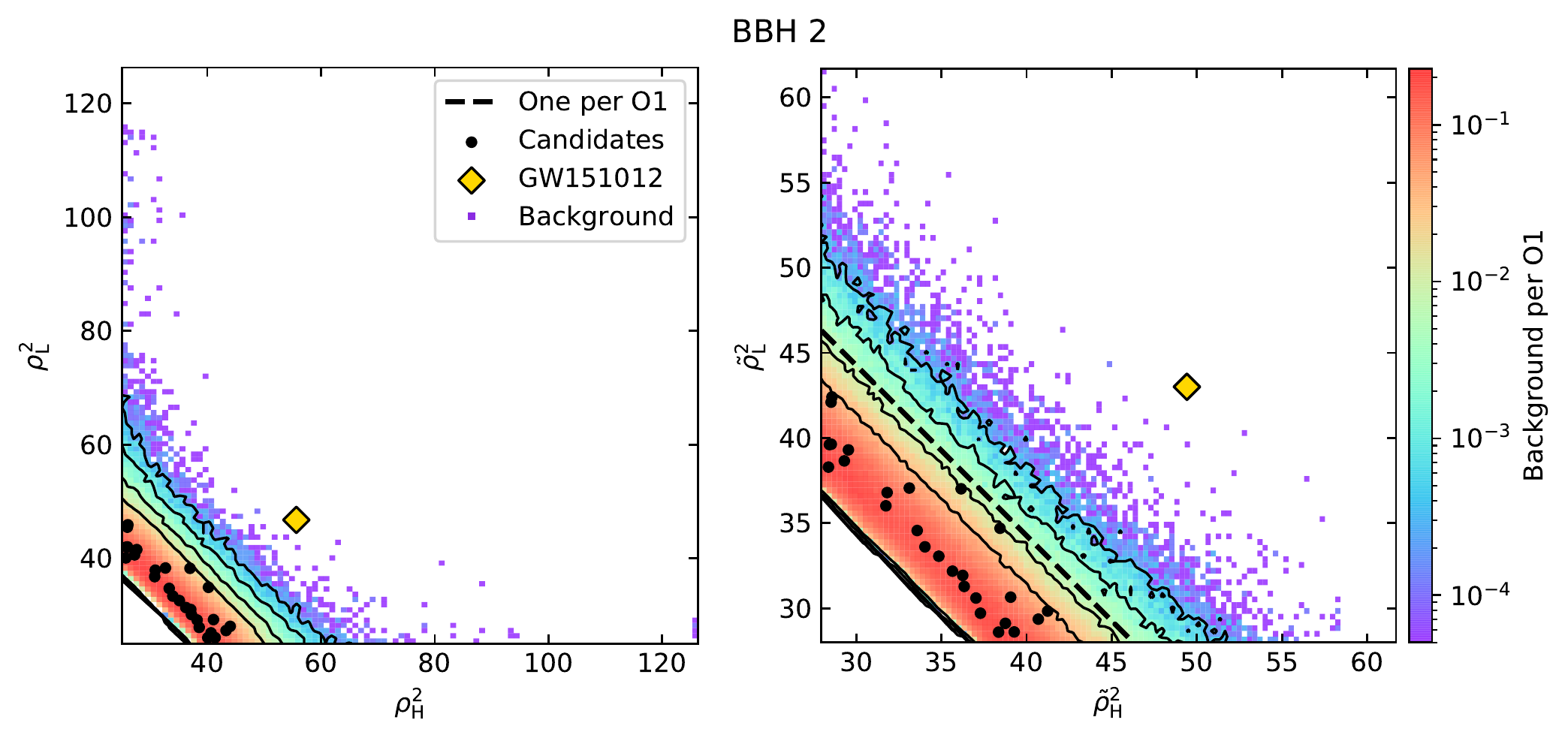}
    
    \includegraphics[width=\linewidth]{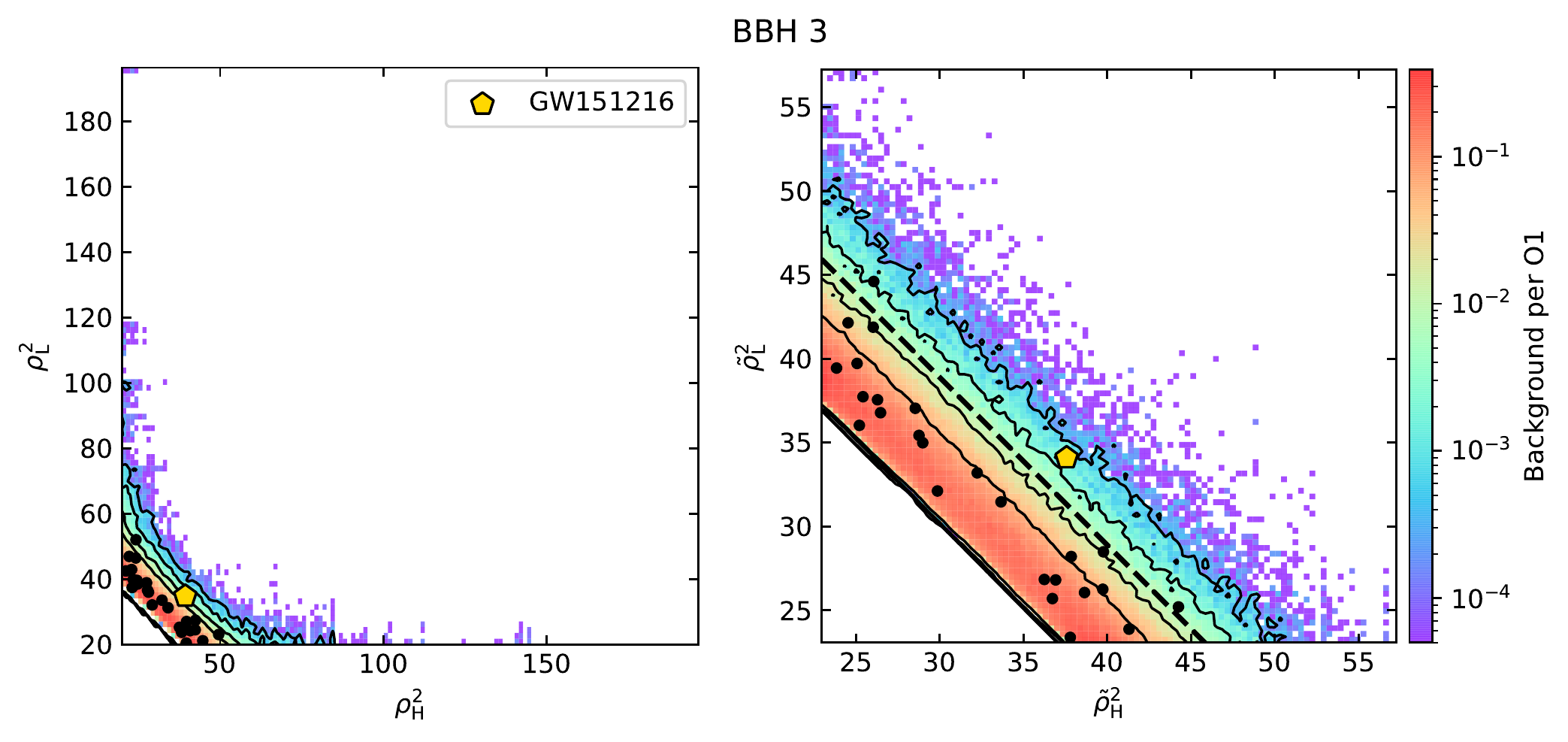}
    \caption{\textit{Left panels:} Two dimensional histogram of the ${\rm SNR}^2=\rho^2$ of the background for the \texttt{BBH 2} (top) and \texttt{BBH 3} (bottom) banks obtained by shifting the data in time so as to recreate \num{2e4} O1 observing runs. The non-Gaussian glitch tail is clearly visible at high {\rm SNR}.\textit{ Right panels:} similar histogram but using the rank-based score $\tilde \rho^2$. The lines of constant probability are straight (solid contours). We show the line corresponding to one event per O1 for this statistic for each bank. 
    Our sub-threshold candidates in these banks are shown together with GW151012 and GW151216.
    GW150914 is too far to the upper right to be included in this histograms.}
    \label{fig:2d-hist-snr}
\end{figure*}

When constructing a statistic to rank events an important part is $P(\rho_{\rm H}^2,\rho_{\rm L}^2 \mid H_0)$, the probability of obtaining a trigger with squared SNRs $(\rho^2_{\rm H},\rho^2_{\rm L})$ in each detector under the null hypothesis $H_0$. Under the assumption that the noise in both detectors is independent, 
\begin{align}
    P(\rho_{\rm H}^2,\rho_{\rm L}^2 \mid H_0)=P(\rho_{\rm H}^2 \mid H_0) P(\rho_{\rm L}^2 \mid H_0)\,.
\end{align} 

If the noise in each detector was Gaussian, 
\begin{align}
    \log P(\rho \mid H_0) = - \rho^2/2 + {\rm const}
\end{align} and 
\begin{align}
    \log P(\rho_{\rm H},\rho_{\rm L} \mid H_0) = -(\rho_{\rm H}^2 +\rho_{\rm L}^2)/2 + {\rm const}.
\end{align} 
Under this assumption it is optimal to use $\rho_{\rm H}^2 +\rho_{\rm L}^2$ to rank candidate events. Unfortunately this is an invalid assumption for two reasons: firstly, even for Gaussian noise, at high ${\rm SNR}$ the maximization over templates, phase and arrival time leads to 
\begin{align}
    \log P(\rho \mid H_0) = -\rho^2/2 + c \log(\rho) + {\rm const},
\end{align}
where the constant $c$ depends on the bank dimension. However, in practice this is a minor correction, the more substantial problem is the non-Gaussian tail of the noise, the so-called glitches. In the high-SNR limit $P(\rho \mid H_0)$ is much larger than the Gaussian prediction. 

The non-Gaussian tail in the $\rho$ distribution has an important consequence when combining the scores of multiple detectors. If we were simply to use $\rho_{\rm H}^2+\rho_{\rm L}^2$ as a score, we would be ranking coincidences in which the trigger in one of the detectors is coming from this non-Gaussian tail, as we would be misjudging its probability by many orders of magnitude. 

To correct this problem we empirically determine $\log [P(\rho_i \mid H_0)]$ for each detector. We do so by taking our triggers and ranking them according to decreasing $\rho_i$ for each detector $i$. We then model \begin{align}
    P(\rho_i^2 \mid H_0) \propto {\rm Rank}(\rho_i^2),
\end{align} 
which is a good approximation for distributions with exponential or polynomial tails. We denote 
\begin{align}
    \tilde{\rho}_i^2 = -2\log P(\rho_i^2 \mid H_0).
\end{align} 
Assuming independence, we can use 
\begin{align}
    \tilde{\rho}^2 = -2\log P(\rho_{\rm H}^2,\rho_{\rm L}^2 \mid H_0) = \tilde{\rho}_{\rm H}^2 + \tilde{\rho}_{\rm L}^2
\end{align} as a robust approximation of the optimal score. In principle, a parametric model for the probability density might outperform the rank estimate, but practical reasons as too few surviving glitches made such estimates prone to fine tuning. Moreover, at the high SNR parts of the distribution, single-detector glitches find background in many timeslides, which makes it problematic to estimate the uncertainty in any such procedure. 
For this reason, and to maintain simplicity, we chose to use the rank function as a proxy for the single detector trigger probability distribution function.

Figure~\ref{fig:rank_vs_snr} shows the relation between $\rho$ and our new Rank-based score $\tilde \rho$ for both LIGO detectors and triggers in bank \texttt{BBH 2}. This mapping is dependent on the bank as the prevalence of non-Gaussian glitch triggers is very different as one changes the length of the templates, i.e., the target chirp mass of the bank. $\tilde \rho$ and $\rho$ agree at low values (only differ by a conventional additive constant), but as $\rho$ increases, $\tilde \rho$ saturates due to the tail in the distribution of triggers. 

In Figure~\ref{fig:2d-hist-snr} we show the two-dimensional histogram of the background obtained by adding \num{20000} unphysical time shifts between detectors to the O1 LIGO data (so as to recreate an equivalent of \num{20000} O1 observing runs) for banks \texttt{BBH 2} and \texttt{BBH 3}. In the left panels we show the distribution of background triggers using $\rho$ as the score. The tail of non-Gaussian glitches is clearly visible leading to an overproduction of triggers where the SNR in one detector is much larger than in the other. On the right panels we show the distribution of the same triggers but now using our rank score to bin them. The lines of constant probability are now straight. Our sub-threshold candidates in these banks are shown together with GW151012, which is a clear outlier, and with GW151216.

For reference, in Figure~\ref{fig:2d-hist-snr} we show the line corresponding to a false alarm rate of one event per O1 observing run based on this statistic. For example, for \texttt{BBH 2} this corresponds to $\tilde \rho_{\rm H}^2\sim\tilde \rho_{\rm L}^2\sim 37$ if divided evenly among both detectors. Figure~\ref{fig:rank_vs_snr} shows that for this threshold {\rm SNR} values the relation between $\rho$ and $\tilde \rho$ is still linear. This demonstrates that although very visible in the histograms, at the detection limit the background is still dominated by the Gaussian part of the noise. The presence of the non-Gaussian glitches does not significantly overproduce the background at the detection threshold. It is also important to note that when we demand that the parameters of the events in both detectors are consistent, according to our so-called coherent score described in the next section, many of these outlier events are heavily down-weighted. 

\subsection{Coherent Score}
\label{sec:coherent}

In this section we further improve the statistic used to rank candidates by exploiting the information encapsulated in the relative phases, amplitudes and arrival times to the different detectors.
We begin with the standard expression:
\begin{equation}
    \max_{T}\frac{P(\rho_{\rm H}^2,\rho_{\rm L}^2,\Delta t, \Delta \phi, t \mid H_1(T))}{P(\rho_{\rm H}^2,\rho_{\rm L}^2,\Delta t, \Delta \phi, t \mid H_0)},
\end{equation}
where $T$ is a template in the continuous template bank.
Because the maximization procedure on $T$ is done incoherently, and prior to the application of all these terms, we will drop it from the notation.
Note that in principle we should have maximized the full expression, but for practical reasons we decided to do the maximization prior to the coherent analysis. In favor of this approximation stands the fact that to linear order, the phase and time shifts are built to be orthogonal to the template identity~\cite{templatebankpaper}, so the template's fine optimization is expected to preserve the $\phi$ and $\delta t$ of a candidate to high accuracy.
We further develop this expression using Bayes rule (and using some basic independence arguments):
\begin{equation}
    \begin{split}
        P(\rho_{\rm H}^2,\rho_{\rm L}^2,\Delta t, \Delta \phi, t \mid H_1) &= P\big(\rho_{\rm H}^2, \rho_{\rm L}^2, \Delta\phi,\Delta t \,\big|\, n_{\rm H}/n_{\rm L}, H_1 \big) \\ 
        &\quad \times P\big(t \,\big|\, H_1, n_{\rm H}^2(t) + n_{\rm L}^2(t)\big) \\
        P(\rho_{\rm H}^2,\rho_{\rm L}^2,\Delta t, \Delta \phi, t \mid H_0) &= P\big(\rho_{\rm H}^2,\rho_{\rm L}^2 \,\big|\, H_0\big) P\big(\Delta \phi,\Delta t \,\big|\, H_0\big),
    \end{split}
\end{equation}
where $n_i$ is the momentary response of detector $i$ computed from the measured PSD, PSD drift correction and the ovelap of the waveform with holes using the data of detector $i$.
$\Delta \phi$ is the difference between detectors in overlap phase of matched filtering the best-fit $T$ with the data.
$\Delta t$  is the difference in arrival time of the maximum score between the detectors.
$P(\rho_{\rm H}^2,\rho_{\rm L}^2 \mid H_0)$ was computed using the ranking approximation detailed in Section \ref{sec:incoherent}.

$P(\Delta \phi,\Delta t \mid H_0)$ is taken to be the uniform distribution by symmetry. Here we note that in principle, $P(\rho_i \mid t, H_0)$ can be non-uniform, if there are bad times where glitches conglomerate. Also, glitches could have a waveform model that prefers a particular phase for a particular template. We currently choose not to introduce these complications (other than the bad times veto applied in Sec. \ref{sec:vetoes}).

$P\big(\rho^2_{\rm H}, \rho^2_{\rm L}, \Delta\phi,\Delta t \,\big | \,n_{\rm H}/n_{\rm L}, H_1\big)$ 
is measured by drawing samples that are uniformly distributed in volume out to a distance where the expected value of the SNR is four, calculating the detector response, and adding noise with the standard complex normal distribution. Out of these samples, we have created a binned histogram of the observed meaningful values $\Delta t, \Delta\phi, \rho_{\rm H}^2, \rho_{\rm L}^2$; the probability of an observed configuration given the signal hypothesis is proportional to the histogram's occupancy.
The same number of samples is used for all values of $n_{\rm H}/n_{\rm L}$ so that the pipeline's preference for detecting events with equal response between the detectors could be evaluated.
This is very similar to the coherent score used in \cite{CoherentScore}.

The term
\begin{equation}
    P\big(t \,\big|\, H_1, n_{\rm H}^2(t) + n_{\rm L}^2(t)\big)\propto (n_{\rm H}^2 + n_{\rm L}^2)^{3/2}
\end{equation}
reflects the changes in sensitivity in the detector as a function of time. Including it allows to analyze different segments of data with very different sensitivities, including multiple runs together (say O1 and O2) while maintaining a consistent detection bar, down-weighting the significance of spurious events from less sensitive detector times.
One important note is that once we include this term, the FAR does not have units of inverse time, but units of inverse volume time.


\begin{figure}
    \centering
    \includegraphics[width=\linewidth]{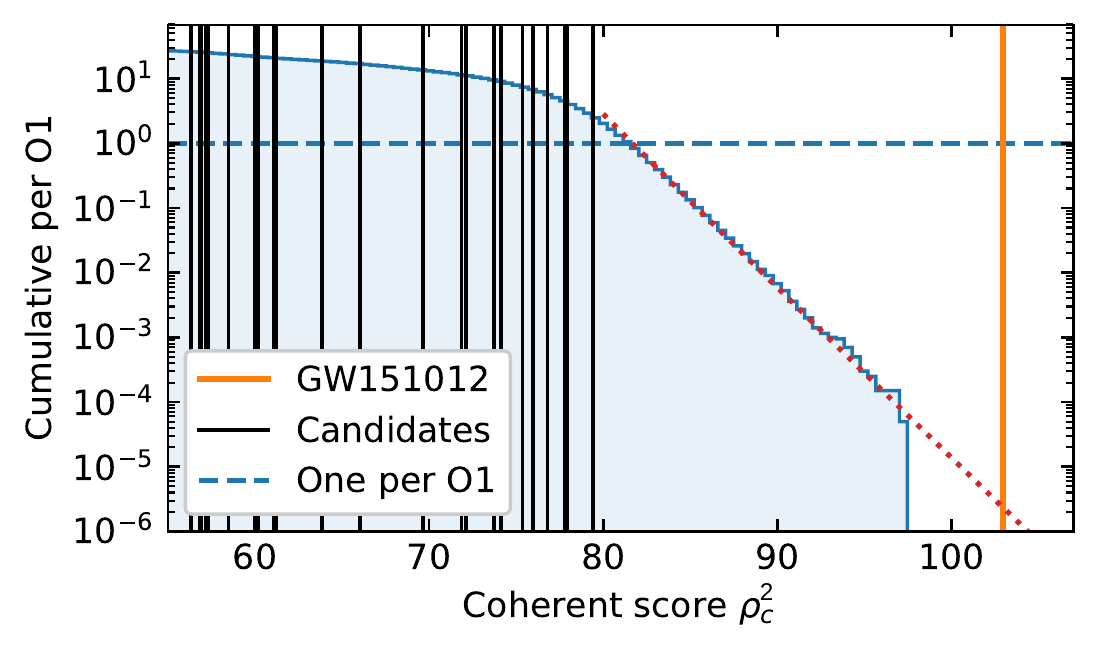}
    \caption{Significance assessment of GW151012. In blue, the cumulative histogram of the coherent scores of background events in bank \texttt{BBH 2} is presented. The flattening at low values is an artifact of the threshold used while collecting background triggers. GW151012 is clearly detected with high significance. We show that its FAR is smaller than 1 in $\num{2e4}$ O1 observing runs. Extrapolation of the background distribution yields a FAR of roughly one in \num{5e5} O1. We note that at this low rate, many more time slides are required for exact assessment of the FAR}
    \label{fig:1d-hist-coh}
\end{figure}

\subsection{Determination of FAR} \label{sec:pastro}

We combine the two detectors in different time-slides with unphysical shifts between $-\SI{1000}{\second}$ and $\SI{1000}{\second}$ in jumps of $\SI{0.1}{\second}$ to obtain an empirical measurement of the inverse false alarm rate of up to $\num{2e4}$ observing runs.
To these unphysical shifts we apply all stages detailed above, exactly as we do the zero-lag data.
Because the optimization and veto stages are computationally expensive, we cannot operate them on all trigger pairs for all time-slide shifts. We ensure that any trigger that has potential of entering the background distribution with an inverse FAR that is better than one per observing run is vetoed, optimized and ranked coherently.

\subsection{Determination of the probability of a source being of astrophysical origin}
While the FAR is largely agnostic of the astrophysical rates (beyond the use of the model in constructing the detection statistic) and is objectively and accurately measurable through time-slides, it is hard to convert to an assessment of the astrophysical origin of a particular event.
Such an assessment depends both on the exact (potentially multidimensional) noise probability density at the event's location (contrast with the one dimensional cumulative probability density the FAR depends on) and the exact probability density given the astrophysical model, including the unknown rate (also as a function of physical parameters).
Essentially, if all exact details in the model were known, the probability of an event being of astrophysical origin would be exactly computable, but in the presence of rate uncertainties, especially when considering the rate as a function of physical parameters, the determination of $p_{\rm astro}$ may be dominated by rate uncertainties and astrophysical prejudice.
Nevertheless, the objectivity of $p_{\rm astro}$ to ranking functions and its immunity to the existence of the few last glitches that are left after our heavy vetoing are compelling, and we therefore proceed in computing it.

To do that, we strictly assume all templates inside a bank are equally probable (even though parameter dependant rate differences probably exist).
We further assume that the background probability density is uniform in time and phase, an assumption we find is extremely good when the SNR value is in the region where the Gaussian noise is dominant.

We then compute the rate at which we observe such an event in coincidence between the two detectors:
\begin{equation}
    \begin{split}
        \mathcal{R}{({\rm event} \mid H_0)} &= \mathcal{R}_{\rm bg}P( \Delta t,\Delta\phi, \rho^2_{\rm H},\rho^2_{\rm L} \mid H_0) \\
        &= \mathcal{R}_{\rm bg}\frac{P(\rho^2_{\rm H} \mid H_0) P(\rho^2_{\rm L} \mid H_0)}{2\pi T},
\end{split}
\end{equation}
where $T$ is the allowed physical time shift between the detector, and $P(\rho^2_{\rm H} \mid H_0),P(\rho^2_{\rm L} \mid H_0)$ were fit using
\begin{equation}
    P\big(\rho_i^2 \,\big|\, H_0\big)=\left(\alpha_i + \beta_i \rho_i^2\right) e^{-\rho_i^2 / 2}.
\end{equation}
$\alpha_i$ and $\beta_i$ are fit to the background computed from time-slides in the region close to the $(\rho^2_{\rm H},\rho^2_{\rm L})$ combination of the event. We find this approximation robust in all cases where the event is close to the detection threshold and when the difference between $\rho^2_{\rm H}$ and $\rho^2_{\rm L}$ is not big.

We then compute the rate ratio
\begin{equation}
    W = \frac{\mathcal{R}({\rm event}\mid H_1)}{\mathcal{R}_{>100}} = \frac{P(\Delta t,\Delta\phi, \rho^2_{\rm H},\rho^2_{\rm L} \mid H_1)}{P(\rho^2_{\rm H}+\rho^2_{\rm L} > 100 \mid H_1)}
\end{equation}
using the table constructed in Section \ref{sec:coherent}.
Here, $\mathcal{R}_{>100} = \mathcal{R}\big(\rho^2_{\rm H}+\rho^2_{\rm L} > 100 \,\big|\, H_1, n_{\rm H}, n_{\rm L}\big)$ is the astrophysical rate of detecting gravitational wave mergers\textit{ in the event's bank}, with the detector sensitivity at the time of the event.
Because $\mathcal{R}_{>100}$ can be easily estimated and updated using a list of known astrophysical events, it is assumed to be known.
We then provide the estimate for the event's astrophysical origin to be:

\begin{equation}
\begin{split}
    p_{\rm astro}({\rm event})
    &= \frac{P{({\rm event} \mid H_1)}}{P{({\rm event} \mid H_0}) + P{({\rm event} \mid H_1)}} \\
    &= \frac{\mathcal{R}_{>100} \frac{W}{\mathcal{R}{({\rm event} \mid H_0)}}}{1 + \mathcal{R}_{>100}\frac{W}{\mathcal{R}{({\rm event} \mid H_0)}}}\,.
\end{split}
\end{equation}
For ease of future interpretation of the results, we report in Section \ref{sec:search_results} both $W/\mathcal{R}{({\rm event} \mid H_0)}$ and the computed $p_{\rm astro}$ using our best knowledge of $\mathcal{R}_{>100}$ at the time of writing.

\section{Results of the BBH search}\label{sec:search_results}

Here we report all the signals and sub-threshold candidates found in the search. We report the FAR in units of ``O1" to reflect the fact that there was a volumetric correction factor in the coherent score. If we assume the sensitivity of the first observing run to be roughly constant, then the ``O1" unit can be converted to roughly 46 days, the effective coincident time we used in the analysis (that has some variation across banks due to differences in the data flagging thresholds).
There was no background trigger with a better coherent score than GW150914, GW151012 and GW151226 in their respective banks, so we obtain only an upper limit on the FAR of $1/(\num{20000}\,\text{O1})$ for all of these events, with an effective $p_{\rm astro}=1$ for all of them. We report their recovered squared SNR for each detector. 
We further found an additional event, GW151216, with a FAR of $1/(\num{52}\,\text{O1})$, reported in greater detail in a companion paper~\cite{GW151216}.
These and two additional sub-threshold candidates with FAR of approximately 1/O1 are reported in Table \ref{tab:signalsFound}.

\begin{table*}
    \centering
    \caption{Events and subthreshold candidates in all of the binary black hole banks.}
    \begin{tabular}{|c|c|c|c|c|c|c|c|c|c|}
        \tabline
        \tabline
         Name & Bank & $\mathcal M (M_\odot)$ & GPS time\footnote{Times are given as the linear-free times, that is, the times corresponding to when the waveforms generated by the bank where orthogonal to the time shift component given the fiducial PSD.} & $\rho_{\rm H}^2$ & $\rho_{\rm L}^2$ & ${\rm FAR}^{-1}$(O1)\footnote{The false alarm rates (FAR) given are computed within each bank. The inverse false alarm rate is given in terms of ``O1" to reflect the volumetric weighting of events using the momentary detector sensitivity.
    Under the approximation of constant sensitivity of the detectors during the observing runs, the unit ``O1" corresponds to roughly $46$ days.} & $\frac{W}{\mathcal{R}(\text{event}\mid H_0)}$ (days) & $\mathcal{R}_{>100} ({\rm days}^{-1})$ & $p_{\rm astro}$  \\
         \tabline
         GW151226 &\texttt{BBH 1} & 9.74 & 1135136350.585 & 120.0 & 52.1 & $>\num{20000}$ & --\textsuperscript{\ref{footnote:noway}} & -- & 1\footnote{We found no credible way of computing the probability density of the background distribution at these high SNRs\label{footnote:noway}.}\\
         GW151012 &\texttt{BBH 2} & 18 & 1128678900.428 & 55.66 & 46.75 & $>\num{20000}$ & $\num{7e5}$ \textsuperscript{\ref{footnote:extrapolation}} & 0.01 & 0.9998\footnote{Estimating $p_{\rm astro}$ for GW151012 required some extrapolation of the background trigger distribution. \label{footnote:extrapolation}}\\
         GW150914 &\texttt{BBH 3} & 28 & 1126259462.411 & 396.1 & 184.3 & $>\num{20000}$ & --\textsuperscript{\ref{footnote:noway}} & -- & 1\textsuperscript{\ref{footnote:noway}}\\
         GW151216\footnote{A new event we are reporting in a companion paper~\cite{GW151216}.} &\texttt{BBH 3} & $29$ & 1134293073.164 & 39.4 & 34.8 & $52$ & $74 \pm 2$ & 0.033 & 0.71\\
         \tabline
         \tabline
         151231 &\texttt{BBH 3} & 30 & 1135557647.145 & 37.5 & 25.2 & 0.98 & $5.4 \pm 0.4$ & 0.033 & 0.15\\
         151011 &\texttt{BBH 4} & 58 & 1128626886.595 & 24.5 & 39.9 & $1.1$ & $16 \pm 1$ & 0.01 & 0.14\\
         \tabline
         \tabline
    \end{tabular}
    \label{tab:signalsFound}
\end{table*}

\section{Conclusions and Discussion}
\label{sec:conclusion}

In this paper we presented an overview of a new and independent pipeline to analyze the publicly available data from the first observing run of Advanced LIGO. 
We used this pipeline to identify a new gravitational merger event in the O1 data. 
In companion papers we will provide additional details of our techniques and implementation choices and further characterize our search by providing simple estimates of the space-time volume searched as a function of parameters. 


There are several areas for future development and improvements in this pipeline, including precise determination of the merger rate/sensitive volume, analysis of single detector triggers, and triggers with subthreshold candidates in the other detector. 
For future runs, it also remains to incorporate more than two detectors into the ranking of coincident triggers in our pipeline.


\section*{Acknowledgment}

We thank the participants of the JSI-GWPAW 2018 Workshop at the University of Maryland, and the Aspen GWPop conference (2019) for constructive discussions and comments.

This research has made use of data, software and/or web tools obtained from the Gravitational Wave Open Science Center (https://www.gw-openscience.org), a service of LIGO Laboratory, the LIGO Scientific Collaboration and the Virgo Collaboration. LIGO is funded by the U.S. National Science Foundation. Virgo is funded by the French Centre National de Recherche Scientifique (CNRS), the Italian Istituto Nazionale della Fisica Nucleare (INFN) and the Dutch Nikhef, with contributions by Polish and Hungarian institutes.

TV acknowledges support by the Friends of the Institute for Advanced Study. BZ acknowledges the support of The Peter Svennilson Membership
fund. LD acknowledges the support by the Raymond and Beverly Sackler Foundation Fund. MZ is supported by NSF grants AST-1409709,  PHY-1521097 and  PHY-1820775 the Canadian Institute for Advanced Research (CIFAR)
program on Gravity and the Extreme Universe and the Simons Foundation Modern Inflationary Cosmology initiative.

\bibliographystyle{apsrev4-1-etal}
\bibliography{gw}

\end{document}